\documentclass[10pt,onecolumn,titlepage]{IEEEtran}
\usepackage{booktabs}

\usepackage{enumitem}
\usepackage{url}
\usepackage[pdftex]{graphicx}
\usepackage{float}
\usepackage{epstopdf}
\usepackage{amsmath}
\usepackage{amssymb}
\usepackage{pifont}
\usepackage[table,xcdraw]{xcolor}
\usepackage{array}
\usepackage{verse}
\usepackage{longtable}
\usepackage{supertabular}
\usepackage{cite}
\usepackage{moresize}
\usepackage{caption}
\usepackage{subcaption}
\usepackage{color}
\usepackage{pifont}
\usepackage{color,soul}
\usepackage{comment}

\newcommand{\cmark}{\ding{51}}
\newcommand{\xmark}{\ding{55}}

\usepackage{url}

\begin{document}

\title{Big Data For Development: \\\emph{Applications and Techniques}}
\author{\large Anwaar Ali$^{1}$, Junaid Qadir$^{1}$, Raihan ur Rasool$^{1,2}$, Arjuna Sathiaseelan$^{3}$, Andrej Zwitter$^{4}$\\
\normalsize $^{1}$School of EE and CS (SEECS), National University of Sciences and Technology (NUST), Pakistan\\
\normalsize $^{2}$King Faisal University (KFU), Kingdom of Saudi Arabia\\
\normalsize $^{3}$Computer Laboratory, University of Cambridge, United Kingdom\\
\normalsize $^{4}$University of Groningen, Netherlands\\
\normalsize \{13mseeaali, junaid.qadir\}@seecs.edu.pk; rrasool@kfu.edu.sa; arjuna.sathiaseelan@cl.cam.ac.uk; a.zwitter@rug.nl\\}
\date{}
\maketitle

\begin{abstract}
With the explosion of social media sites and proliferation of digital computing devices and Internet access, massive amounts of public data is being generated on a daily basis. Efficient techniques/ algorithms to analyze this massive amount of data can provide near real-time information about emerging trends and provide early warning in case of an imminent emergency (such as the outbreak of a viral disease). In addition, careful mining of these data can reveal many useful indicators of socioeconomic and political events, which can help in establishing effective public policies. The focus of this study is to review the application of big data analytics for the purpose of human development. The emerging ability to use big data techniques for development (BD4D) promises to revolutionalize healthcare, education, and agriculture; facilitate the alleviation of poverty; and help to deal with humanitarian crises and violent conflicts. Besides all the benefits, the large-scale deployment of BD4D is beset with several challenges due to the massive size, fast-changing and diverse nature of big data. The most pressing concerns relate to efficient data acquisition and sharing, establishing of context (e.g., geolocation and time) and veracity of a dataset, and ensuring appropriate privacy. In this study, we provide a review of existing BD4D work to study the impact of big data on the development of society. In addition to reviewing the important works, we also highlight important challenges and open issues.\end{abstract}

\section{Introduction}
\label{sec: intro}


In the modern world we are inundated with data, with companies such as Google and Facebook dealing with petabytes of data \cite{mayer2013big}. Google processes more than $24$ petabytes of data per day, while Facebook, a company founded a decade ago, gets more than $10$ million photos \emph{per hour}. The glut of data, buoyed by fast advancing technology, is increasing exponentially due to increased digitization of all aspects of modern life (using technologies such as the Internet of Things (IoT)\cite{iotmckinsey2015}--which uses sensors, for example in the shape of wearable devices, to provide data related to human activities and different behavioral patterns). It is estimated that we are generating  $2.5$ quintillion bytes per day (we note here that a quintillion bytes, or an exabyte, is equal to $10^{18}$ bytes) \cite{siegel2013predictive}.

The presence of ``big data'', or this massive amount of increasing data, offers both an opportunity as well as a challenge to researchers. A lot of progress has been made in developing the capability to process, store, and analyze big data: In addition to the big data computing capability (in terms of processing and storing big data in a distributed fashion on a cluster of computers \cite{barroso2013datacenter}), the rapid advances in using intelligent data analytics techniques---drawn from the emerging areas of artificial intelligence (AI) and machine learning (ML)---provide the ability to process massive amounts of diverse unstructured data that is now being generated daily to extract valuable \emph{actionable} knowledge. This provides a great opportunity to researchers to use this data for developing useful knowledge and insights \cite{hilbert2013big}.



From the perspective of big data for development (BD4D), an important quandary is gaining access to important people-related data, which is often in the exclusive access of the government in the form of paper documents. Fortunately the emerging trend known as ``\textit{open data}'', which promotes open public sharing of data from various public and private sector entities in searchable and machine-readable formats is a boon for BD4D research. Governments worldwide (e.g., in USA  \cite{usgov_data,hoffmann2012data} and the UK \cite{ukgov_data}) are increasingly adopting open data projects to fuel innovation and transparency. In addition, open source platforms have been developed that facilitate the creation and gathering of digital data from mobile platforms (e.g., the Open Data Kit \cite{hartung2010open}). While open data can be rightly regarded as a subset of all the available big data: the nuance is in the liquidity of big data \cite{manyika2013open}. Open data also promotes a culture of creativity and public wellbeing as is evident by different hackathons that are being organized to tap the potential of open data in terms of useful mobile applications (e.g., the local government of Rio de Janeiro has created the Rio Operation Center \cite{rio_online,wordbank2014bd} aimed at harnessesing the power of technology and big data to run the city effectively in terms of transport management, natural disaster relief, mass movement and management of slum areas). In a recent report from McKinsey Global Institute \cite{manyika2013open}, the net worth of open data was estimated to be \$3 trillion. In this report, the importance of open data is highlighted for seven particular sectors: education, health, transportation, consumer products, electricity, oil and gas, and consumer finance.

In 2009, the Secretary-General of the United Nations (UN), Ban Ki-moon started the \textit{UN Global Pulse} initiative, with the explicit goal of harnessing big data technology for human development \cite{pulse2012big}. The Global Pulse program is aimed at forming a network of innovation centers, called the \emph{Pulse Labs}, all over the world. Ideally, these Pulse labs will bring together people from different fields of life together to make use of the free and open source computing methods/ software toolkits to analyze data to help the development and humanitarian operations especially in the developing countries. In \cite{kirkpatrick2013big}, Kirkpatrick, the director of the UN Global Pulse innovation initiative, presents the case for deploying big data techniques and analytics in the field of human development. It is highlighted that data---especially from mobile phone and social media---can be utilized in fighting hunger, disaster and poverty. This report talks about ``data philanthropy'' where the companies, whose businesses revolve around data, can collaborate with the UN in predicting imminent humanitarian crises and help take possible steps to avoid situations that can lead to disasters. The report also discusses the issues and challenges faced by the UN in terms of data access, user privacy and the integration of big data techniques into the various UN humanitarian systems.

The aim of this paper is to answer the important question: how can we harness the big data technologies to transform and revolutionize the developing world? Towards this end, we will review the applications of big data techniques in the context of development and thereby highlight the potential development areas that can benefit from big data technology. We believe that consistent with the huge impact of big data on all other facets of modern society \cite{mayer2013big,siegel2013predictive}, big data also has an immense potential for the field of international human development.  We will consider questions such as:

\begin{enumerate}

\item How to access and use all of the data that is present out there on the isolated servers of the companies and organizations for the development purposes?

\item What particular areas of development can benefit from big data?

\item What are some of the well-known techniques for big data analytics that can be applied in the BD4D context?
\end{enumerate}


\vspace{1mm}
\emph{Contributions of this Paper}: Despite the great potential of BD4D, the research field of BD4D is only nascent. In this study, we have chosen not to approach the problem of BD4D only from a technological viewpoint, since development is a nuanced subject, we have chosen to adopt a multidisciplinary vantage point (integrating technology, economics, social and development sciences). For this paper, we have reviewed existing research literature, official documents, online projects, blogs and technical reports related directly or indirectly to BD4D. Apart from highlighting the immense potential of BD4D, our work also identifies some of the associated challenges and potential lurking harms that must be understood and countered. Our paper is distinct from existing survey papers \cite{hilbert2013big} \cite{wordbank2014bd} in that apart from highlighting the particular development areas that can benefit from big data, we also discuss various techniques for big data analytics, while also describing open issues and directions for future work.


\vspace{1mm}
\emph{Organization of this Paper}: In Section \ref{sec: BD_techniques_tools}, we present necessary background related to different techniques that are used to analyze, store and process big data. In Section \ref{sec: dev_areas}, we present the broad spectrum of areas where big data can play a role in human development. In Section \ref{sec: bd_ana4d}, we discuss big data analytical techniques in the perspective of mobile, living and visual analytics and link these techniques to various human development opportunities. In Section \ref{sec: challenges}, we present a discussion on the challenges of using big data for development and identify open issues and future work. Finally, in Section \ref{sec: conclusion} we conclude our study with a discussion and our stance related to the revolutionary and transformative power of big data in modern society.

\section{Background: Big Data Techniques}
\label{sec: BD_techniques_tools}

Modern datasets, or the \emph{big data}, differ from traditional datasets in 3 V's: volume, velocity and variety. In today's age huge volumes of data is being generated at huge pace (or velocity) and the numerous sources of data give vast variety to it. All of this data, if harnessed intelligently, can truly realize the notion of the \emph{information age} \cite{hilbert2013big}. Actionable information can be gathered from the data after performing intelligent processing and analytics on the available data. The techniques (specially related to machine learning) in order to gather, store, process and analyze this vast amount of data are the subject matter of this section. We also try to link this discussion, and different examples considered here to explain various concepts, to the humanitarian development. The aim of this section is to provide readers with a brief background and related work of the relevant techniques to help them understand their applications when discussed in the perspective of humanitarian development.

\subsection{Machine Learning}
\label{subsec: ml}

Machine learning (ML), a sub-field of artificial intelligence (AI), focuses on the task of enabling computational systems to learn from data about how to perform a desired task automatically. Machine learning has many applications including decision making, forecasting or predicting and it is a key enabling technology in the deployment of data mining and big data techniques in the diverse fields of healthcare, science, engineering, business and finance. Broadly speaking, ML tasks can be categorized into the following major types:

\vspace{2mm}
\subsubsection{Supervised Learning}
\label{subsubsec: super_learn}

In this class of ML, the learning task is to generalize from a training set, which is labeled by a ``supervisor'' to contain information about the class of an example, so that predictions can be made about new, yet unseen, examples. If the output (or prediction) belongs to a continuous set of values then such a problem is called \emph{regression}, while if the output assumes discrete values then the problem is called \emph{classification}. In the following we briefly present a few classification techniques.

\begin{itemize}

\vspace{1mm}
\item \emph{Naive Bayes Classifiers} are based on Baye's Theorem that assume independence among features given a class \cite{rish2001empirical}. These has been widely used for the Internet traffic classification: e.g., naive Bayesian classification of the Internet traffic \cite{moore2005internet}.

\vspace{1mm}
\item \emph{Decision Trees (DT)} define a popularly used intuitive method that can be used for learning and predicting about target features both for quantitative target attributes as well as nominal target attributes. Although, DT do not always perform very competitively, their main advantage is their intuitive interpretation which is crucial even network operators have to analyze and interpret the classification method and results.

\vspace{1mm}
\item \emph{Support Vector Machines (SVM)} is a widely used supervised learning technique that is remarkable for being practical and theoretically sound, simultaneously. The approach of SVM is rooted in the field of statistical learning theory, and is systematic: e.g., training a SVM has a unique solution (since it involves optimization of a concave function).

\end{itemize}

\vspace{2mm}
\subsubsection{Unsupervised Learning Techniques}
\label{subsubsec: unsuper_learn}

The basic method in unsupervised learning is \emph{clustering}. In clustering, the learning task is to categorize, without requiring a labeled training set, examples into `clusters' on the basis of perceived similarity. This clustering is used to find the groups of inputs which have similarity in their characteristics. Intuitively, clustering is akin to unsupervised classification: while classification in supervised learning assumed the availability of a correctly labeled training set, the unsupervised task of clustering seeks to identify the structure of input data directly.

\vspace{2mm}
\subsubsection{Reinforcement Learning}
\label{subsubsec: rein_learn}

This is a \emph{reward/ punishment} based ML technique. In this technique a learner, based on an input received, performs some action, potentially affecting the environment around it. This action is then rewarded or punished. The nature of the mapping from the actions taken by the learner to rewards/ punishments, in general, is probabilistic in nature. The eventual goal of a learner is to discover such an optimal mapping (or policy), from its actions to the rewards/ punishments, so that the average long-term reward is maximized.

\vspace{2mm}
\subsubsection{Deep Learning}
\label{subsubsec: deep_learning}

Deep learning (DL) is an ML technique that comprises deep and complex architectures \cite{schmidhuber2015deep, bengio2009learning}. These architectures consist of multiple processing layers, each capable of generating non-linear response corresponding to the data input. These layers consist of various small processers running in parallel to process the data provided. These processors are called \emph{neurons}. DL has proved to be efficient in pattern recognition, image and natural language processing \cite{collobert2008unified}. DL finds its applications in very broad spectrum of applications ranging from healthcare to the fashion industry \cite{dl_mit_fashion}, with many key technology giants like Google, IBM and Facebook deploying DL techniques to create intelligent products.

\vspace{2mm}
\subsubsection{Association Rule Learning}
\label{subsubsec: assoc_learn}

 It is a method for discovering interesting relations between variables in large databases. In this, we seek to learn about \emph{associations} between the features present in examples. Unlike classification (supervised learning), which strictly and discretely tells the class of an example, relations or associations among various variables in an example database are considered in association rule learning. We take an example case mentioned in \cite{witten2005data} where a weather dataset is considered. The usual classification problem would be to tell whether, based on the values of given weather features or attributes (like temperature, outlook and wind conditions) in the dataset, a game would be played or not. If, however, we consider association learning perspective then (instead of always telling about the status of the game) different rules among different features or variables can also be considered. As a example a rule can be established that if the outlook is sunny and the game is being played then the day is going to be non windy. This type of learning technique can be particularly important for farmers in planning their activities for the best possible crop productions.

\vspace{2mm}
\subsubsection{Numeric Prediction}
\label{subsubsec: num_pred}

In \emph{numeric prediction}, we are not interested in predicting the discrete class (or category) to which the example belongs, but the numeric quantity associated with it. As an example consider, once again, the weather dataset mentioned to explain the association learning. Now consider the classification problem where instead of predicting whether (based on the given features) a game will be played or not a numeric quantity, e.g., how long (in minutes) a game is likely to be played, is predicted as an output \cite{witten2005data}. The same scenario, again, can be of importance to a farmer where a numeric quantity such as time, how long, or how much rain will fall on a particular day can be predicted.

\subsection{Data Mining, Knowledge Discovery, and Data Science}
\label{subsec: dm_kd_ds}

Data mining usually refers to automated pattern discovery and prediction from large volumes of data using ML techniques \cite{witten2005data}. Data mining can also be used to refer to online analytical processing (OLAP) or SQL queries that entails retrospectively searching a large database for a specific query. OLAP queries, also known as decision-support queries, are typically complex expensive queries that take a long time and touch large amounts of data. The process of extracting useful information or knowledge from the structured/ unstructured data and databases (relational and non-relational), using data mining and ML techniques, is called \emph{knowledge discovery}, sometimes collectively called \emph{KDD (knowledge discovery in databases)}. This knowledge can be in the form of brief and concise visual reports, a predicted value or a model of a larger data generating system \cite{fayyad1996data}. Data science is an inderdisciplinary field in which different KDD techniques and processes are studied. Next, we briefly describe the trend of non-relational databases to store unstructured data followed by an introduction to predictive analytics that helps in knowledge discovery from the huge volumes of structured/ unstructured data.

\vspace{2mm}
\subsubsection{New Trend in Database Technology: NoSQL}
\label{subsubsec: db_tech}

With the advent of big data and Web $2.0$ we now have a huge amount of unstructured data such as word documents, email, blog posts, social- and multimedia data. This unstructured data is different from the structured data in that it can not be stored in an organized fashion in the conventional relational databases. In order to store and access unstructured data, a different approach and techniques are required. NoSQL (or non-relational) databases have been developed for the same purpose \cite{leavitt2010will}. Companies like Amazon (Dynamo \cite{decandia2007dynamo}) and Google (Bigtable \cite{chang2008bigtable}) adopt this approach for storing and accessing their data. The main advantage, besides storing unstructured data, is that these NoSQL databases are distributed and hence easily scalable, fast and flexible (as compared to their \emph{relational} counterpart). One of the concerns in using NoSQL datases, though, is that they usually do not inherently support the ACID (atomicity, consistency, integrity and durability) set, as supported by the relational databases. One has to manually program these functionality into one's NoSQL database.

\vspace{2mm}
\subsubsection{Predictive Analytics}
\label{subsubsec: pred_ana}

Predictive analytics refers to a technology that aims to provide a competitive advantage by predicting some future occurrences or behavior (using data mining and ML techniques) based on past experience (in the form of collected data). Predictive analytics encompasses data science, machine learning, predictive and statistical modeling and outputs empirical predictions based on given input empirical data \cite{shmueli2010predictive}. The underlying premise is that future can be predicted on the basis of the past experience. Although, this premise matches our every day intuition, it is problematic philosophically due to the \emph{problem of induction} which asks the question: `can the future be predicted on the basis of the past?'. Notwithstanding the objections, it has been borne out in practice that although we can not deterministically tell the future, in many cases, we can improve our decision making by probabilistically reasoning about future predicted outcomes---though care must be taken to also consider the proverbial `black swan' that may appear unexpectedly to ambush our predictions. Predictive analytics finds its application in various humanitarian development fields ranging from healthcare to education. As we advance through the text we discuss the applications of predictive analytics in more detail in the upcoming sections.

\subsection{Crowdsourcing and Big Data}
\label{subsec: crowd_bd}

Crowdsourcing is different from \emph{outsourcing}. In crowdsourcing, the nuance is, a task or a job \emph{is} outsourced but not to a designated professional or organization but to general public in the from of an open call \cite{howeblog_crowsourcing}. Crowdsourcing is a technique that can be deployed to gather data from various sources such as text messages, social media updates, blogs, etc. This data can then be harmonized and analyzed in mapping disaster struck regions and to further enable the commencement of search operations. This technique helped during the $2010$ Haiti earthquake \cite{meier2015digital}. Crowdsourcing, based on social media, is discussed in \cite{gao2011harnessing} in terms of the opportunities that it provides for disaster relief and the challenges that are being faced during this process.

\subsection{Internet of Things}
\label{subsec: iot}

Internet of things (IoT) is a new trendy field fueled by the hype in big data, emergence of network science \cite{barabasi2015networkscience}, proliferation of digital communication devices and ubiquitous Internet access to common population. A technical report by McKinsey Global Institute \cite{iotmckinsey2015}, presents the potential of IoT in terms of economic value. According to the study conducted in \cite{iotmckinsey2015}, if all the challenges are overcome, the IoT has a potential to create $3\$$--$11\$$ trillion USD worth of economic value. In IoT, different sensors and actuators are connected via a network to various computing systems providing data for actionable knowledge. In this way IoT, big data and network science are all related. \emph{Interoperability}, harmony of data from one system with another, is a potential challenge in the way of IoT expansion. IoT finds its application in healthcare monitoring systems. Data from wearable body sensory devices and hospital health care databases, if made interoperable, could help doctors to make more efficient decisions in diagnosing and monitoring chronic diseases. Similarly, with the help of ML techniques and predictive analytics, data that is fed in real-time to computing systems by sensors and actuators can be utilized to revolutionize the maintenance tasks in industries with a significant reduction in the breakdowns of parts and system downtimes.

\section{Big Data for Development: Development Areas}
\label{sec: dev_areas}

In this section, we will highlight some of the major development areas in which BD4D is applicable. We will first explore the role of BD4D in times of natural disasters and political crises. Besides these humanitarian emergencies, we study how BD4D can be used in the fields of agriculture, healthcare, education and in the alleviation of poverty and hunger. To accompany the information contained in this section, we have also presented a tabulated summary of projects pertaining to these different development areas in Table I.

\begin{table*}[!ht]
\centering
\scriptsize
\caption{Big Data Projects for Different Development Areas and Tools for Big Data Analytics}
\label{tab: bdProjects}
\begin{tabular}{p{3.5cm}p{1cm}p{2.25cm}p{1.25cm}p{8.50cm}}

\hline
\cellcolor[HTML]{EFEFEF}\textbf{\textit{Project}} &
\cellcolor[HTML]{EFEFEF}\textbf{\textit{Reference}} &
\cellcolor[HTML]{EFEFEF}\textbf{\textit{Type}} &
\cellcolor[HTML]{EFEFEF}\textbf{\textit{Open Data}} &
\cellcolor[HTML]{EFEFEF}\textbf{\textit{Description}}  \\
\hline

\\
\multicolumn{5}{l}{\textbf{\cellcolor[HTML]{EFEFEF} \textit{Humanitarian Emergencies}}}\\ \\

Ushahidi & \cite{ushahidi_online} & Online Service & \xmark & Crowdsourcing platform, merging data from various sources for various humanitarian emergencies. \\
Digital Humanitarian Network & \cite{dhn_online} & Online Service & \xmark & Network of IT volunteers to leverage the technology to fight human crises. \\
Trace the Face & \cite{tracetheface_online} & Non-profit & \xmark & Crowdsourced online platform to help find separated migrants. \\
GNUcoop & \cite{gnucoop_online} & Online Service & \xmark & Network of IT professional to deploy information and communication technologies for development (ICT4D). \\
Open Street Map & \cite{osm_online} & Non-profit & \xmark & Real-time, online and crowdsourced map service to map a natural crisis of a region. \\
Pakistan Body Count & \cite{pbc_online} & Online Analysis & \xmark & Makes use of publicly available data to map the casualties caused by suicide bombing and drone attacks in Pakistan. \\
Services Advisor (UNHCR) & \cite{serviceprovider_map} & Web App & \xmark & Real-time, interactive map services for migrant to find aid agencies' operations and locations quickly. \\

\\
\multicolumn{5}{l}{\textbf{\cellcolor[HTML]{EFEFEF} \textit{Healthcare}}}\\ \\

Personal Genome Project (PGP) & \cite{pgp_online} & Non-profit & \cmark & Publicly shared genomic data for research purposes. \\
$1000$ Genomes Project & \cite{1000genomes_online, 10002010map} & Non-profit & \cmark & Human genome sequencing to study the relationship between phenotypes and genotypes. \\
Google Flu Trends & \cite{gflutrends_online} & Non-profit & \cmark & Inactive now, but provides data about flu and dengue trends for different regions. \\
Data.gov (Health) & \cite{usgov_data_health} & Governmental & \cmark & Open health data, tools and applications from the US Government. \\
UNGP (Health Projects) & \cite{ungp_projects_online} & Non-profit & \xmark & Provides health case studies over different regions of the world. \\
Health Data (The World Bank) & \cite{twb_data_health} & Non-profit & \cmark & Publicly available health data from The World Bank's projects and studies. \\
Individualized Health Initiative (JHU) & \cite{hopkins_individual_online} & Academic & \xmark & Research initiative to promote individualized healthcare by combining and analyzing patients' data from various sources. \\
Human Connectome Project & \cite{humanconnectomeproject, humanconnectome_online} & Non-profit & \cmark & Accurate mapping of human connectome. \\

\\

\multicolumn{5}{l}{\textbf{\cellcolor[HTML]{EFEFEF} \textit{Education}}}\\ \\

PSLC Data Shop & \cite{koedinger2010data, datashop_online} & Non-profit & \cmark & Open data repository of learning data for educational data mining. \\
Educational Data Mining & \cite{idms_online} & Non-profit & \cmark & Community dedicated for R\&D based on learning data mining of different education data.  \\
Data.gov (Education) & \cite{usgov_data_edus} & Governmental & \cmark & Open data, tools and apps related to education at all levels from the US Government. \\
Education Data (The World Bank) & \cite{twb_data_edu} & Non-profit & \cmark & Open data from The World Bank's projects and studies related to education. \\
Alltuition & \cite{alltuition_online} & Online Service & NA & Makes use of open education data to provide students' with best possible educational financial opportunities. \\
Simple Tuition & \cite{simpletuition_online} & Online Service & NA & Provides information related to different financial opportunities provides by the educational institutes. \\
Mastery Connect & \cite{mastery_online} & Service Provider & NA & Tool provides teachers with the real-time understanding level of each student of his/her class. \\
Knewton & \cite{knewton_online} & Service Provider & NA & Adaptive tool for to enable individualized learning and teaching. \\
ThinkCERCA & \cite{thinkcerca_online} & Service Provider & NA & Personalized online teaching tool, enable teachers to design and teach according to the changing standards. \\
UNGP (Education Projects) & \cite{ungp_projects_online} & Non-profit & \xmark & Educational case studies over different regions of the world. \\

\\

\multicolumn{5}{l}{\textbf{\cellcolor[HTML]{EFEFEF} \textit{Miscellaneous}}}\\ \\

Billion Prices Project & \cite{bbp_online} & Academic & \cmark & Data from many online retailers are analyzed for near real-time economic research. \\
Predictify.me & \cite{predictify_online} & Service Provider & \xmark & Data collection and predictive analytcs for business growth and market understanding. \\
Be Data Driven & \cite{bdd_online} & Service Provider & \xmark & Development of products for various organization based on their data. \\
First Mile Geo & \cite{fmg_online} & Service Provider & \cmark & A platform for data collection, visualization and analysis. \\
Data.gov & \cite{usgov_data} & Governmental & \cmark & Provides open data for different field, e.g., health, education, agriculture etc. \\
UNGP & \cite{ungp_online} & Non-profit & \xmark & Dedicated to deploy technology and especially big data for development. \\
The World Bank (Data) & \cite{twb_data_online} & Non-profit & \cmark & Provides open data from its various studies and projects for different fields. \\

\\
\multicolumn{5}{l}{\textbf{\cellcolor[HTML]{EFEFEF} \textit{Tools for Big Data Analytics}}}\\ \\

Open Data Kit (ODK) & \cite{odk_online} & Open Source Tool & NA & Provides online tools for data collection and analysis. \\
RapidMiner & \cite{rapidminer_online} & Open Source Tool & NA & Predictive analytics platform. \\
Hadoop & \cite{hadoop_online} & Open Source Tool & NA & Open source tool for distributed computations on large amounts of data. \\
Weka & \cite{hall2009weka, weka_online} & Open Source Tool & NA & Open source tool to deploy machine learning algorithms for analytics on big data. \\
Elastic Map Reduce (Amazon) & \cite{emr_amazon_online} & Processing Tool & NA & Data processing service for large amounts of data. \\

\\

\hline
\end{tabular}
\end{table*}

\subsection{Humanitarian Emergencies}
\label{subsec: hum_emer}

In this section, we will present two case studies related to natural disasters and political crises through which we will highlight the important role that can be played by big data. Different issues concerning the acquisition, storage and sharing of data under these emergencies are also considered.

\vspace{2mm}
\subsubsection{Natural Disasters}
\label{subsubsec: nat_dis}

In his book \cite{meier2015digital}, Meier talks about the important and crucial role that the analysis of big data can play when a natural disaster strikes a part of the world. When an earthquake hit Haiti in $2010$, after this incident the community of online users played a very significant role to fight this disaster. Through \emph{crowdsourcing}, a real-time image of the situation, or a \emph{crisis map}, became clear. Big data techniques from the fields of AI and ML were deployed to find meaning in massive and fast-changing online data comprising of tweets and short message service (SMS), which was generated after the disaster. The author calls members of this community the \emph{digital humanitarians}. This book introduces the concept of big \emph{crisis} data. In its very nature it is not different from the usual big data except that it is created especially in times of crisis and disaster. By employing analytical techniques with the help of ML tools and methods, useful and actionable information can be extracted from this data. The author also outlines various potential challenges and harms that lurk behind the usage of this big crisis data. The most important among these is the credibility of data. With the ubiquitous online connectivity and proliferation of digital communication devices, a fake dataset or trend can be easily generated. The author talks about efficient AI and ML techniques to verify these data.

\vspace{2mm}
\subsubsection{Migrant Crisis}
\label{subsubsec: mig_crisis}

As we write this paper, the ongoing political unrest in Syria, which started in $2011$, worsens day after day. This situation has displaced a large number of people internally and a staggering number outside the country. The fleeing of people from the troubled areas, leaving their own homes and finding shelter elsewhere, has resulted in mass movement of population---the magnitude of which has not been observed since the end of World War II. Syria's immediate neighbors---in particular, Lebanon and Jordan---and many countries in Europe are seeing a huge influx of people in search of shelter, better and less troubled lives. In this type of scenario, it is a challenge for the humanitarian organizations to operate efficiently especially in the troubled and war-torn areas. Two major challenges are faced by the helping organizations: (i) ensuring that the right regions get the right type of assistance in time; and (ii) ensuring the coordination within and among organizations during such times. This is important to avoid chaos and mismanagement. In both of these cases, data has a vital role to play.

 The United Nations Refugee Agency (UNHCR) in collaboration with non-profit volunteer organizations \cite{afinitybridge_online, peacegeeks_online, gnucoop_online}, formed by people with IT skills to deploy technology for humanitarian purposes, developed an online map, called Services Advisor \cite{serviceprovider_map}, fueled by the organizational data from UNHCR's \textit{ActivityInfo} \cite{activityinfo_online}. This map is interactive and can be accessed from a variety of desktop and mobile phone browsers. A user can zoom in to his/ her nearest location and get to know about the number of different organizations, their function, operating times and capacity. In this fashion accurate and real-time information is made available to the refugees so that they can get help quickly and avoid long queues and other similar inconveniences. Through ActivityInfo portal, different aid agencies can crowdsource their information, related to location, services and number of people they are serving, so that coordination can be established among all the working agencies in troubled areas.

A number of projects related to the use of big data for human development and dealing with humanitarian emergencies  are listed in Table I. An important concern with most of the BD4D projects dealing with humanitarian emergencies is that they essential spring to action after the crisis has taken a huge toll. The real promise of BD4D is to use \emph{predictive analytics} \cite{siegel2013predictive} to avoid or mitigate such humanitarian emergencies before they can strike their toll. It is worth noting that similar predictive strategies are being deployed in most other businesses---e.g., the retail giant Amazon predicts what a user would like based on its past behavior and purchases. To sum up, for the future the immense amount of data, especially from the projects that has already been started during this crisis, must be utilized to prevent such situations in the first place.
Towards this end, there needs to be research on development of models based on data corresponding to various social and political indicators. As an illustrative example, this sort of predictive analytics when done right could have been afforded the ability to develop models that would have predicted the ongoing Syrian `migrant crisis' \cite{wired_refugee}.

\subsection{Hunger, Food and Agriculture}
\label{subsec: hunger}

Kshetri in \cite{kshetri2014emerging} surveys recent research literature and official reports/ documents to study the factors that help enable the use of big data techniques for development purposes along with the inhibitors in the way of this process. The importance of modern data sources, e.g., social media and cell phone data, is highlighted. In terms of skills, monetary capacity to \emph{afford} data, and sometimes cultural and industrial norms for utilizing modern technology result in nonuniform diffusion of technological innovations and trends throughout the world. This paper presents a case study for agriculture to discuss the opportunities and challenges for deploying big data techniques for the development of farmers.

In the developing countries, the farmers are often less informed about the soil conditions, extreme changes in the weather patterns, plantation, topography and access to markets \cite{kshetri2014emerging, wordbank2014bd, connectedalliance_online}. Data collected from different sensors, satellite imagery and field experts can be analysed and predictive models can be formed. Based on these models the most relevant information can then be sent using cellular network to individual farmers.

\subsection{Healthcare}
\label{subsec: healthcare}

Big data analytics in healthcare is bringing a huge cultural change in the way conventional medical diagnosis and treatment operates. Big data can revolutionalize medical diagnosis by integrating data gathered from various medical records of a patient, as well as real-time wearable sensors, to analyze and diagnose the patient's current health status and provide an early warning sign if the health of a patient is on a dangerous track. Doing this helps in taking preventing measurements to diagnose and treat a potentially harmful disease during early stages. In terms of making treatment more efficient and convenient, it is possible for a person having a smart phone to access medical service providers via a healthcare app \cite{researchkit_apple_nyt} to obtain quick and more personalized response from the convenience of one's home.

Adopting a modern technologically-driven approach, combining both medical and data sciences, has great implications for the medical practice. Currently most of the patients' data is being stored in electronic form on different databases of different medical service providers. The challenge today is that all of this data, though in the electronic form, sits on different locations in the form of \emph{``fragments''} \cite{elhauge2010fragmentation}, that by itself provides an incomplete picture to the corresponding medical-care provider. If the challenging issue of integrating these fragments can be resolved, there is a healthy prospect of \emph{democratization} of health information \cite{lohr_nyt_healthcaredemocratization}, through which the  study of disease can flourish by combining medical science and data science. This integration can further be expanded to cover a whole country to construct a \emph{Learning Health System (LHS)} \cite{friedman2014toward} in which the faculties from policymaking, medical-care, engineering and technology are merged together to analyze and fight diseases rapidly and more accurately. This system has the potential to create an environment where research and clinical practice are not performed separately; rather new research and analysis are directly applied to patients in near real-time fashion. Expanding this concept further and covering the whole world could provide valuable information about the current status of any country's health and early warning signs of imminent viral epidemic outbreaks. This can be only the first step in this process; the next step is to provide relevant medical assistance, immunization vaccines and related preventive measures to a specific region of the world.

In existing work, practical systems have been built that have used big data technology for building an early warning system for a potential epidemic breakout. As an example, Pervaiz et al. presented a study of comparative analysis of the performance of different algorithms that are deployed on Google Flu Trends \cite{gflutrends_online} to detect an early warning sign of a potential epidemic breakout \cite{pervaiz2012flubreaks}. A number of other health-related BD4D projects are summarized in Table \ref{tab: bdProjects}. Among these projects, the ones related to human genome are very important. Sequencing of a human genome creates massive amounts of data that is crucial in understanding the origin and dynamics of various diseases.

\subsection{Education}
\label{subsec: education}

The field of education is making a transition to digital era with the use of physical textbooks waning and digital versions of study material gaining more popularity. Education is one of the fields that has greatly benefited from the big data analytics \cite{picciano2012evolution}. The conventional pedagogical practices, students' learning and study habits, and the way whole educational system is being designed and run are seeing  revolutionary changes.

In particular, the practice on online learning and blended learning is gaining popularity. In blended learning, online teaching, learning and assessments are combined with the conventional pedagogical approach.


There are two important interrelated big data related developments in education: learning analytics and educational data mining. \textit{Learning analytics (LA)} is an emerging cross-disciplinary field that combines data analytics and learning, thus bringing researchers together from various fields such as computer science, data science and social science \cite{siemens2012learning}. In this field, research is carried out for various purposes that include, but are not limited to, predictive analysis, social network and sentiment analysis, personalized learning, and better curriculum designs. \textit{Educational data mining (EDM)}, like LA, is also an emerging and related field. In EDM, data mining and ML techniques are applied to the data representing the student's interaction with the digital and online educational system---which can be easily stored in massive open online courses (MOOCs) and online tests---to help the students better learn. With the proliferation of digital devices, and the increased consumption of the Internet and social media sites, every Internet user leaves behind a \emph{data trail} \cite{datafloq_education}, which can be exploited to learn and understand a person's behavior. In the context of online learning, recording and mining the student's interaction with the learning system has the potential of revealing interesting insights that can be exploited to optimize the student's learning experience. Many vendors are producing data driven products for educators with user-friendly interfaces to bridge research outputs and real practices \cite{siemens2012learning}. There is a dedicated community named \emph{International Educational Data Mining Society} \cite{idms_online}, whose sole purpose is to provide platform for people and researchers to publish and develop effective techniques and solutions based on data mining for effective learning and teaching.

Through LA and EDM, the conventional teaching and learning methods are being modified. Different data-driven products are available for the teachers to design tests for students and in turn the data, related to student behavior and level of understanding, is collected. Different aspects are analyzed during and at the end of such assessment processes. Data, such as related to students' answers to different questions, how long a student took to answer a specific question, how often a student has to click other links to understand the question statements can be collected and a finalized, mostly visual, analytical report is presented to the teacher. This renders a teacher to quickly  pinpoint students who are struggling with specific topics or questions. In this way an \emph{individualized treatment} can be given to such students to address their particular problems so that they can be brought up to the mark.

In a similar manner data from different teachers can be analyzed together. This can give insights in terms of which teacher has the greatest mental harmony with what kind of students. As an example, a teacher might be struggling with shy students while the same type of students show better results with a different teacher. In this way an early and informed \emph{intervention} can be performed so that such cases can be resolved in time. Three of such projects from different vendors can be seen in references \cite{mastery_online, knewton_online} and \cite{thinkcerca_online}.

\section{Big Data Analytics for Development}
\label{sec: bd_ana4d}

\subsection{Mobile Analytics}
\label{subsec: mob_ana}

Mobile analytics is the application of big data techniques to the massive amounts of data that mobile companies gather about their users in terms of call volume, calling pattern, and location. This data contains a wealth of information that can be very useful for research, planning and development (the use of such information also poses many privacy and ethical use challenges). The field of mobile big data analytics focuses on analyzing cell-phone data to provide insights that can be used to drive value-added services. For example ``call-detail-records'' (CDR) analysis maintained by mobile service providers can be used for gathering socioeconomic information. Mobile Data Challenge \cite{laurila2012mobile} by Nokia research was one of the projects aimed at gathering and utilizing mobile phone data for research purposes. The paper \cite{laurila2012mobile}, describes the project details, its purpose and the research methodology. Around $200$ smart-phone users volunteered their mobile phone data in Switzerland for the purpose of this research project.  The collected CDR data was multimodal; rich with the information related to mobility, communication and interaction patterns. This data was further utilized, after ensuring privacy, for the research purpose and was made publicly available so that worldwide research collaborations, to analyze this data, can be made. Technical (strict and secure data storage and data anonymization techniques) and agreement based approaches (between volunteers and researchers) were adopted to ensure the privacy of the volunteers during both the data collection and making it available to the research community. If privacy and other issues related to big data are taken care of then these projects are very important in enabling the research efforts to explore the immense potential that the big data has to impact the future of technology.



Social network analysis is an important field of research where cell-phone data provides valuable information and useful insights. A study \cite{candia2008uncovering} utilizes cell-phone data for social network analysis. In this study the data is analyzed in terms of space and time. Through this \emph{spatiotemporal} analysis of cellphone activity, mean collective behavior of humans are analyzed and special focus is given to the occurrence and spread of anomalous behavior through a social network. Concepts and tools from the \emph{standard percolation theory} \cite{stauffer1994introduction}, which deals with the pattern and behavior of clusters in a given graph, are deployed to map and quantify the spread of anomalous patterns in a space at a given time.  In terms of time, this analysis can be extended by taking consecutive slices of time. This shows the spread, pattern, and decay of the anomalous behavior in time. Overall, this kind of analysis provides a very detailed and accurate picture of emergency situations, which can be in terms of political turmoil or spread of an epidemic. Analysis of individual call activity pattern is also studied in the same work. This analysis provides information about the mobility of people, which would further help in planning effective transportation strategies.

Another study that combines network science and big data is conducted in \cite{onnela2007structure}, for big data driven social-network analysis. In this study, a huge dataset of mobile phone interactions between individuals of a certain region is utilized to construct a social network to study the relation between the topology of this network and the \emph{tie strengths} between users that comprise this social network. Tie strengths are the measure of link weights between individual users. A (non-directional) link exists between two users if there is, at least, one reciprocated call between them. Call duration determines the numerical value of the weights of these links. As a result, the user call data helps to reproduce the structure of a social network where users are nodes and they are connected with different tie strengths. An interesting result of this study is that the removal of strong ties has minimal effect on the integrity of the structure of this social network. However, if the weak links are removed, the structure of the social network collapses. Another interesting finding is that ties with intermediate strengths are basically more useful in spreading information in a social network as opposed to both the strong and weak ties. These insights provide  wonderful opportunities in understanding the dynamics of a social network, and to plan effective policies for the population of these networks. Medical awareness campaigns, as an example, can be specially designed so that the links with intermediate tie strengths are targeted to spread the information effectively throughout the network.

Mobile analytics can be used for a number of developmental purposes such as urban planning, transport engineering, analysis of social dynamics of a group of people, and even epidemic control. We will briefly consider a few representative example use cases: (i) In \cite{frias2012computing}, the authors have used the CDR information to develop a \textit{CenCell} tool \cite{frias2012computing} to aid governments and policy makers in computing reasonably accurate and affordable census maps by approximating census information using anonymized CDRs using supervised and unsupervised classification techniques; (ii)  In \cite{frias2012estimation}, the authors have used mobile analytics on CDR to model commuting patterns to help characterize the mobility of the human population and thereby generate commuting matrices; (iii) Finally, in \cite{frias2011agent}, the authors have proposed \textit{AlertImpact} as a method to analyze the evolution of an epidemic under various policies by performing mobile analytics on anonymyized CDRs. By using AlertImpact to analyze anonymized CDRs collected during the H1N1 outbreak in Mexico, the authors were able to document 10\% reduction in peak number of individual virus infections  due to the government mandates.

\subsection{Living Analytics}
\label{subsec: liv_ana}

\begin{table*}[!ht]
\centering
\scriptsize
\caption{Data Sources and Application Areas of Big Data Analytics for Development}
\label{tab: ana_app_sources}
\begin{tabular}{p{3cm}p{3.5cm}p{7cm}p{3cm}}

\hline
\cellcolor[HTML]{EFEFEF}\textbf{\textit{Data Type}} &
\cellcolor[HTML]{EFEFEF}\textbf{\textit{Medium}} &
\cellcolor[HTML]{EFEFEF}\textbf{\textit{Application Area(s)}} &
\cellcolor[HTML]{EFEFEF}\textbf{\textit{References\newline(Literature and Projects)}}  \\
\hline

\\
\multicolumn{4}{l}{\textbf{\cellcolor[HTML]{EFEFEF} \textit{Mobile Analytics}}}\\ \\

Call Detail Records & Cell Phones & Social Network Analysis, Population Mobility Patterns, Transportation System Planning, Awareness Campaigns, Mobile App. Usage Patterns & \cite{laurila2012mobile, onnela2007structure, candia2008uncovering, stauffer1994introduction, eagle2006reality,frias2012computing,frias2012estimation,frias2011agent} \\

\\

\multicolumn{4}{l}{\textbf{\cellcolor[HTML]{EFEFEF} \textit{Living Analytics}}}\\ \\

Tweets and Comments & Social Media Sites & Social Network Analysis, Sentiment Analysis & \cite{wordbank2014bd,ushahidi_online, meier2015digital} \\

Search Queries & The Internet & Epidemic Prediction, Employment Status & \cite{wordbank2014bd, gflutrends_online, pervaiz2012flubreaks} \\

Click History & The Internet & Recommender Systems, User Behavior & \cite{wordbank2014bd}\\

Text & The Internet & Cultural Changes, Policy Effectiveness & \cite{wordbank2014bd} \\

Financial Data & Credit/ Debit Card Transactions & Fraud Detection, Marketing & \cite{wordbank2014bd} \\

Personal Health Data & Wearables & Healthcare & \cite{wordbank2014bd,iotmckinsey2015} \\

\\

\multicolumn{4}{l}{\textbf{\cellcolor[HTML]{EFEFEF} \textit{Visual Analytics}}}\\ \\

Images & Satellites & Mass Movement, Surveillance, Disease Control, Transportation & \cite{wordbank2014bd} \\

Climate Variables, Temperature, Pollutant Levels & Sensors & Weather Forecasting, Pollution Control, Urban Planning & \cite{steele2010beautiful, rio_online} \\

Geo-Location & Cell Phones & Mobility Patterns, Social Network Analysis & \cite{eagle2006reality, nyt_visualization} \\

\\

\hline
\end{tabular}
\end{table*}

Living analytics is a big-data-driven interdisciplinary field of research that incorporates expertise from a number of disciplines including computer science, network science, social science, and statistics. Living analytics is related to the study of social and behavioral patterns of individuals and societal groups. Like other fields, social science has also advanced through the recent development in big data technology: the field of \emph{computational social science} is inherently based around using the advances in storage and computing capabilities to process readily available big data for advancing our understanding of social science \cite{lazer2009life}. Conventional social science techniques, which are mainly based on questionnaires and surveys, suffer from bias, incompleteness or sometimes inaccurate and scarce information. Modern techniques where data from devices, specially cell-phones and other digital communication devices, are collected and different models are formed to study the structure and dynamics of a social network either on individual or collective levels are in contrast with the conventional methods. Intelligent techniques are being devised and deployed to mine useful data from the massive datasets gleaned from cell-phones and other digital devices.

Computational social science represents a paradigm shift from traditional social science in many profound ways: instead of manually gathering data using one-time personalized surveys, the availability of digital data (such as GPS location, email logs) can provide more dynamic visibility into human behavior both at the level of an individual and at the level of society.

The important thing, as mentioned in \cite{lazer2009life}, is to address all the barriers in the way of the development of computational social science. This work \cite{lazer2009life} describes two types of challenges: One is the approach and the other one is related to infrastructure. The research, or academic, approach in computational social science should be to access and gather data through secure, channelized mechanisms. There should be centralized data storage facilities under the supervision of technological savvy personnel that understand the threat of security breach in the data. This is opposed to the approach where data is distributed under people with varying technological skills and security protocols.

Mining of data to gain insights into the social patterns of an individual or a group of individuals falls into the realm of \emph{reality mining}. A reality mining related research from MIT researchers was published in $2006$ \cite{eagle2006reality}. In this study, $100$ cell-phones were distributed among students and faculty members. Data related to proximity, location and device usage was collected from these devices over a period of nine months. These were collected mainly from the cell tower logging, providing user location, and Bluetooth, providing information related to proximity. Techniques from information theory were deployed to model individual behaviors, like device usage patterns and mobility. Similarly collective models, representing friendship, acquaintance and ethnographic networks were developed based on the information provided by this data. This work also studies the behavioral patterns on an organizational level. These insights are important in predicting possible future behavior, both on individual and collective levels. This helps in developing efficient planning in times of crisis.



Outside the realm of research and academics, it is important to devise policies and technologies aimed at collecting data related to human behavior while at the same time protecting user privacy and user comfort. This calls for effective \emph{human machine interfaces (HMIs)} \cite{pantic2007human}. The survey \cite{pantic2007human} published in $2007$ discussed the relationship between humans and a computer. It is argued, in this article, that if computing is to be used effectively and solely for humans, in terms of comfort and assistance, then a paradigm shift is needed in the way interaction occurs between a human and a computer. The emphasis is put on \emph{human-centered} technological designs instead of conventional \emph{computer-centered} designs. This is to be done so that computers can \emph{sense} humans in the same \emph{natural} way humans interact with each other. This interaction can be in terms of voice input, facial expressions, or even physiological condition of a human body, where sensing is often done without a subject being consciously aware of it. Different questions are raised in this article, and related literature pool has been reviewed to address these. These questions are related to \emph{what} data is being collected by \emph{which} means and \emph{why} is it being collected and \emph{how} a proper response can be generated based on the information collected by addressing all of these questions? Addressing these questions provides us with the information about in what format the data is collected? what computing or sensory interface was used? In which context (time and location) the data was collected? And based on this information what is the most appropriate response to address the user needs or queries.

\subsection{Visual Analytics}
\label{subsec: vis_ana}

Visual analytics is an interesting branch of big data exploration in which the aim is to support the science of analytical reasoning through interactive visual interfaces. Through information visualization, large amounts of quantitative data can be shown in a limited space \cite{tufte1983visual}. As mentioned in \cite{keim2002information}, in visual analytics there might not be much a priori information known about the data or even about the data exploration goals. In information (or data) exploration the goals are steered and fine-tuned during the process of exploration by human interaction. Visual analytics has the power to quickly convey the essence of a massive dataset to a user as contrast to automatic data mining and machine learning tools, which require more technological soundness and knowledge. As an example, as it quite often happens these days, the viral trends on any one of the social media sites, e.g. Twitter or Facebook, can provide one with a good idea of the trend if this outbreak is shown in an animated time-lapsed video. One can track the origin and hubs responsible for the spread of the virus. Through these principles, combined with the concepts from the network science, the outbreak of biological viruses can also be analyzed or even prevented beforehand.

 In this section, we will focus on four approaches to visualization and visual analytics: data maps, time-series, space-time narratives, and relational graphics. We briefly describe these approaches, as discussed by Tufte in \cite{tufte1983visual}, and link them to humanitarian development:

\vspace{1mm}
\subsubsection{Data maps} This type of visualization is usually a cross of cartography and statistical information. In data maps a region of interest is considered and a specific variable under-consideration is analyzed over the spatial dimensions of this region. Quoting an example from the book \cite{tufte1983visual}, a map of the USA is considered and death-rates due to different types of cancers are shown all over the map. This statistical information, that is the death rates, are shown by coloring different counties of the US according to the death-rates' statistical information. A user can easily locate which counties suffered the most due to cancer and which suffered the least by consulting the color scheme provided with this data map. Provided with the additional information related to the socioeconomic norms of a county one can investigate the reasons, hubs and links in the spread of the disease. This type of information can thus be very useful in healthcare and specially in epidemiology to prevent a potential outbreak of a viral disease.

\vspace{1mm}
\subsubsection{Time series} In this type of representation, the growth, development, decay or the general trend of a variable is presented against the time lapse. Time can be in various resolutions, ranging from seconds to centuries. The rise and fall of stock market over a period of time, temperature variation of a specific region and, as discussed in \cite{eagle2006reality}, a user's device usage over different times of a day are all examples of time series. Time series are important to analyze trends that arise during a specified period (e.g., dengue mosquitoes that mostly bite during dawn and dusk and during specific months of a year). This type of information provides disaster management authorities a prior information to take preventive steps to avoid large number of casualties during crisis times.

\vspace{1mm}
\subsubsection{Space-time narrative} In the time-series, as discussed above, if the variable of \emph{space} is also introduced into the analysis then, as a result, we have a multivariate data for visualization. Tufte in \cite{eagle2006reality} mentions that to attain excellence in presenting the information in graphics, one almost always has to deal with a multivariate information. The visualization enables one to easily understand the substance provided in the graphic while being less aware of the complex multivariate nature of the underlying data. An example of this narrative provided in \cite{eagle2006reality} would adequately describe the importance of space-time visualization in development process. The example is related to environmental pollution where the concentration of different pollutants are observed at different times of a day over different regions of an area. We end up with different time series each corresponding to a specific pollutant. Slices taken from any one of these time series are basically data maps, similar to the ones discussed before. In short, a slice taken from a time series corresponding to a pollutant reveals the levels of this pollutant over different regions of an area at the specific time the slice is taken from. As a result we have the information about at what times a pollutant peaks in its concentration and at which place. This helps us to analyze the dynamics of a particular region at a specific time. As an example carbon mono-oxide (variable one: Pollutant) level peaks during traffic rush hours (variable two: Time) specially at the intersection of large roads (variable three: Space).

\vspace{1mm}
\subsubsection{Relational graphics} In this type of visualization the variables can take any form or type. Like mentioned before, in these types of graphics the relation between two or more quantities is analyzed, which are not necessarily only time and space. An example of this kind of analysis can be number of deaths per million versus cigarette consumption pattern over a range of a spatial region \cite{eagle2006reality}. The variable of time can also be added to this analysis, extending the experiment to extract the changing patterns of deaths because of cigarette consumption over different periods of times. The resulting graphic will show how effective the campaign, against smoking, really is over a period of time by observing the decrease in the deaths in different regions. If, for example, a few regions are showing resistance then the focus can be diverted to this particular area. More variables, like those related to sociopolitical norms, can further be added to pinpoint the troubles and ideally address the issues.

\section{Challenges, Open Issues and Future Work}
\label{sec: challenges}

         In today's age it is quite likely that big data will gain substantial potential and importance in order to shift the paradigm of the conventional humanitarian development process in almost every walk of life. It is, however, not a panacea to all the problems in the modern day world. Just like any other innovation, the wide scale adoption of big data is hindered by many potential challenges. In this section we discuss some of these challenges from two perspectives: technical and ethical. Correspondingly, we describe open issues and future work, which is required to address these challenges.

\subsection{Technical Challenges}
\label{subsec: tech_chal}

There are various technical challenges involved in implementing BD4D. As an example, with the daily production of vast amounts of data, are the processing and storing capabilities scaling proportionally? Below we present and describe a few of the technical challenges:

\begin{itemize}

\vspace{1mm}
 \item Crowdsourcing: We observed the importance of crowdsourcing when we discussed migrant crisis. Social media sites are a rich source of crowdsourced data and many aid agencies rely on the information gathered from these sites. However, there is not an established framework where the agencies can collaborate, and ideally complement each other's efforts and findings. This produces a problem of \emph{double response}. Where two aid agency take same action on the same problem when, if there were coordination between these two, one of the agencies could be operating on a different problem or taking a complementary action for the same problem.

\vspace{1mm}
 \item Bias and Polarization: The personalized content predicted by algorithms based on the past behavior of a user can create polarization. This means that two different users could be getting entirely different search results for a same thing. However, modern deep-learning techniques, which do not entirely rely on the past data, and context aware computing and algorithms can address these issues. As an example Facebook has a dedicated research group by the name of \emph{Facebook's AI Group}. One of the tasks that this group aims to complete is to find meaning in the user posts that is not entirely based on keyword search \cite{emtech_fb_meaning}. A similar venture is by IBM. Cognitive computing \cite{modha2011cognitive}, which is based on deep-learning and brain inspired neural algorithm approaches, will help the development of Watson, a context aware AI based computing system \cite{ibmwatson}.

\vspace{1mm}
\item Data Supply Chain: With all the benefits, policy analysis by utilizing big data is a precarious task. Many potential challenges and perils entail this process \cite{schintler2014big}. Privacy is a major, and largely debated, concern in gathering data from users. During the data gathering process (the \emph{big-data supply chain}) the context and semantics of the data can be altered resulting in faulty and sometimes controversial policies. Present day data sources are also prone to temporal and spatial restrains, due to disparity in worldwide technology proliferation, resulting in a statistical bias, which in turn can result in inefficient policies.

\vspace{1mm}
 \item Technology Usage: The context, specially in the online data collected about students, is very crucial to consider in LA. A problem that arises, while tracking the data-trail left by students online, is that every individual has a different attitude towards the usage of technology. The social network and sentiment analysis should be performed with care so that the students who use the Internet less, or differently, as compared to other students should not be penalized in the data analysis \cite{eynon2013rise}.

\vspace{1mm}
 \item Spatial Problem: Many users update their status with the information related to a crisis sitting, all together, at a different geographical site. This is a challenge in pin pointing an actual place of crisis for which the information was provided at the first place. So, the data gathered from the actual ground based surveys and aerial imagery should be corroborated with these for the effective actions to fight a crisis situation.

\vspace{1mm}
 \item Vulnerability of Connectivity: Although, scientists are working on trust management systems for the verification of the information gathered for an appropriate action: Fraudulent information and entities can still infiltrate the information network. This information can then be treated like normal data and has the potential to diffuse and infect other connected entities of the information network. This vulnerability is primarily caused by the connected nature of information producing and consuming entities, this \emph{vulnerability of connectivity} and cacasding errors/ failures are discussed by Barabasi in his book \cite{barabasi2015networkscience}.

\vspace{1mm}
 \item Interoperability: Big data analytics often include collecting and then merging unstructured data of varying data types. As an example call detail records from cell phone companies, satellite imagery data and face-to-face survey data have to be corroborated together for the better and less-biased analysis. Merging and harmonizing this data for analysis is a challenging task. For effective data analytics a system is needed that could make data streams of potentially different formats homogenous.

\vspace{1mm}
 \item Fragmentation: The challenge of fragmentation is one of the major impediment to large-scale deployment of big data analytics. As an example, a patient might be seeing different specialists for, seemingly, different medical reasons. These specialists, then further, can prescribe different types of clinical tests resulting in different kinds of results.  If, however, some protocol or a system is developed to integrate these fragments together and run analysis on them collectively then a clear and big picture of a patient's current health can be extracted. If the issue of fragmentation is resolved then this can not only speed up the diagnosis process but also has the ability to provide personalized treatment most suitable for the patient under consideration.

\vspace{1mm}
 \item Technology Scaling: In recent times, the technologies of cloud computing and software-defined networking (SDN) have proved very useful for efficiently implementing big data solutions: going forward, more work is needed to ensure that the computing and networking facilities scale to the ever-increasing scale of data \cite{qadir2014sdns}.

\end{itemize}

\subsection{Ethical}
\label{subsec: eth_chal}

Besides all the technology related challenges presented above it is imperative to consider the ethical dimensions of utilization BD4D. Throughout the paper, we have tried to outline, besides all the benefits, the potential challenges and harms incurred by the deployment of big data for development purpose. We saw that privacy is one of the major issues in almost every field where big data analytics are applied. Besides privacy, the challenge of fragmentation is one of the major impediment to large-scale deployment of big data analytics. Besides these well-known issues, there are a few subtle challenges as well: most of which fall into the realm of ethics and abuse of technology. Here we list a few of the challenges faced in the perspective of ethics when dealing with BD4D.

\begin{itemize}
\vspace{1mm}
 \item Privacy: This concern tops the list. As an example, with large amounts of data being collected about individuals, it is of utmost importance that such information should not be abused for any sort of personal or financial gains.

\vspace{1mm}
 \item Digital Divide: This divide \cite{hilbert2011end} is simply the nonuniform diffusion of technological advancement and expertise through out the world. The result of this divide harm nations that lack the infrastructure, economic affordability and \emph{data-savvy} faculty. The digital divide, the well-known issue of privacy, and the control and monopoly of entities exploiting the data are among the important challenges that hinder the wide scale deployment of big data techniques for development.

\vspace{1mm}
 \item Open Data: There are also many possible issues with open data. For greater transparency, it is desirable that government/ development data is openly accessible. However, it is also important to think about who has the right to access, use, link, and repurpose open data (and how much flexibility is desirable, keeping in view various misuse and privacy issues). With the rising use of big data in humanitarian and development aid, governance efforts should focus on ensuring that sensitive information (such as the location of humanitarian actors and IDPs) does not become open, since such data may maliciously be exploited by malevolent actors.
  \end{itemize}

  Finally, the evolution of data science, in itself is a challenge. This is because the field requires expertise and collaboration of people from various fields and disciplines. Interdisciplinary efforts should be encouraged and financially incentivized so that big data can be analyzed with the right perspectives and ethics in place.

\section{Conclusions}
\label{sec: conclusion}

In this paper, we have reviewed the literature focused on using big data techniques for human development (BD4D). Our aim in this paper is to highlight to a broad audience the immense potential of BD4D in a variety of settings including humanitarian emergencies (including disaster response and migrant crisis), agriculture, poverty alleviation, food production, healthcare and education. We have highlighted the various challenges and pitfalls associated with BD4D. We envision that in the future BD4D will play a big role in human development and global prosperity, but to succeed with BD4D, it is imperative that researchers are able to tackle and solve the challenges identified.

\section*{Competing Interests}
The authors declare that they have no competing interests.


\bibliographystyle{ieeetr}
\bibliography{BD4D}

\end{document}